\documentclass[twocolumn,longauthor]{aastex61}

\received{23 January 2018}
\revised{23 February 2018}
\accepted{1 March 2018}

\shorttitle{KIC 8462852 Photometry 1922 to 1991}

\shortauthors{Castelaz and Barker}

\begin{document}

\title{KIC 8462852: Maria Mitchell Observatory Photographic Photometry 1922 to 1991}

\correspondingauthor{Michael Castelaz}
\email{michael.castelaz@brevard.edu}


\author{Michael Castelaz}
\affiliation{Division of Science and Mathematics, Brevard College, Brevard, NC 28712, USA; michael.castelaz@brevard.edu}

\author{Thurburn Barker}
\affiliation{Astronomical Photographic Data Archive, Pisgah Astronomical Research Institute, Rosman, NC 28772, USA; tbarker@pari.edu}




\begin{abstract}

A new study of the long-term photometric behavior of the the unusual star KIC 8462852 (Boyajian's Star) has been carried out using archival photographic plates from 1922--1991 taken at the Maria Mitchell Observatory (MMO).  We find five episodes of sudden, several day, decreases in magnitude occurring in 1935, 1966, 1978, and two in 1980.   Episodes of sudden increase in magnitude appear to occur in 1967 and 1977.  Inspection of archival light curves of KIC 8462852 from two previous studies based on the Harvard and the Sonneberg plate collections finds apparent corresponding events to these observed episodes in the MMO light curve.  Also, a general trend of $0.12\pm0.02$ magnitudes per century decrease is observed in the MMO light curve, significant, but less than the trend of 0.164$\pm$0.013 observed in the Harvard light curve.

\end{abstract}




\section{Introduction} \label{sec:intro} 

KIC 8462852 (Boyajian's star; J2000 20h 06m 15.455s +44$\degr$ 27$\arcmin$ 24.793$\arcsec$) is an F3V star which decreased suddenly in brightness and lasting several days (hereafter referred to as dips) in the Kepler bandpass by 16\% in 2011 and again in 2013 by 21\%, 3\%, and 8\%, in addition to six other dips of 0.5\% and less during the Kepler mission \citep{2016MNRAS.457.3988B}. A series of dips in 2013 were separated by a total of about 50 days. The star dipped again a few percent in May 2017 \citep{2017AAN...579....1W}. Using 800 days of photometry from the All-Sky Automated Survey for Supernovae \citep[ASAS-SN;][]{2014ApJ...788...48S} and 4,000 days of photometry from the All-Sky Automated Survey \citep[ASAS;][]{2002AcA....52..397P}, \cite{2017arXiv170807822S} find two brightening episodes lasting several hundred days and a steady decrease in magnitude of 6.3$\pm$1.4 mmag yr$^{-1}$. \cite{2017arXiv170901732K} reanalyzed the Kepler data and found  22 dips, with two dip events separated by 928.25 days, each lasting $\sim$2 days and dropping 1010$\pm$40 ppm. Furthermore, \cite{2017RNAAS...1...22G} observe U-shaped fading separated in time by 4.4 years. Besides dips and brightening episodes,  \cite{2016ApJ...830L..39M} found cumulative fading of 3\% over the four year Kepler mission which is similar to the dimming observed by \cite{2017arXiv170807822S}. 

Explanations for long term dimming and brightening and dips include models of obscuration that could result from the catastrophic destruction of a planet, large comet fields impacting the star, asteroids, and planetary/massive object transits \citep[for various scenerios see][]{2016MNRAS.457.3988B}.  One transit model includes Trojan-like asteroids and ringed planet model \citep{2018MNRAS.473L..21B} and makes the prediction that large decreases will occur again in the year 2021 based on a $\approx$12 year orbital period of a ringed planet. Another transit model indicates a 4.31 year period attributed to a transit or groups of transits  \citep{2017arXiv171001081S}. The transit model of \cite{2017A&A...600A..86N} show that the light curve can be explained with four massive objects where each massive object is surrounded by a dust cloud and all four objects are in similar eccentric orbits.  \cite{2017MNRAS.468.4399M} develop models of planetary consumption where the dips are due to obscuration by planetary debris from the disruption of a Jupiter-mass planet.  This study also explains the slow long term dimming of KIC 8462852.  \cite{2018arXiv180100732B}  observed 4 dips from 1\% to 4\% beginning in May 2017 until the end of 2017.  Their observations are consistent with obscuration by ordinary dust.  

Internal mechanism models used to describe the variations of KIC 8462852 were originally proposed by \cite{2016ApJ...829L...3W}. \cite{2017ApJ...842L...3F} explains the long and short term variability of KIC 8462852 with a model of star spots where the stellar convective zone stores the heat flux. Dips in the magnitude of KIC 8462852 may be the star storing its radiative flux.  \cite{PhysRevLett.117.261101} modeled the small few tenths of a percent dips in the Kepler light curve of KIC 8462852 as intrinsic transitions, but this model does not explain the deep, sudden dips like those observed in 2011 and 2013.     Clearly, continued observations are needed to constrain the external and intrinsic mechanism models developed to explain the light curve.  

Photometry from archival photographic plates can provide data useful for testing models by searching for long term variations and dips looking back in time many decades.   Photometric studies extending back more than 100 years have been conducted from historic data archived in photographic plate collections.  \cite{2016ApJ...822L..34S} presents a light curve from 1338 Harvard College Observatory plates over the period 1890 to 1989.  The digitization and photometry was provided by the Digital Access to a Sky Century @ Harvard \citep[DASCH;][]{2009ASPC..410..101G,2013PASP..125..857T} project.  The light curve from the Harvard data suggests KIC 8462852 is dimming 0.164 $\pm$ 0.013 magnitudes per century. 

\cite{2016ApJ...825...73H} accessed the Sonneberg Observatory photographic plate collection \citep{1992Stern..68...19B}  to produce light curves from 861 B magnitudes (Sonnberg--pv data) and 397 V magnitudes (Sonneberg--pg data) covering the period from 1934 to 1995. The light curve shows constant magnitude to within 0.03 magnitudes per century, or about a 3\% decrease in brightness, consistent with the ASAS light curve and Kepler data.  The same study also used some plates from the Sternberg Observatory and Pulkovo Observatory plate collections. 

The Astronomical Photographic Data Archive at the Pisgah Astronomical Research Institute \citep[]{2009ASPC..410...70C,2014aspl.conf....4B} contains a set of astronomical photographic plates consistently taken with the same telescope from the Maria Mitchell Observatory \citep[MMO;][]{2009ASPC..410...96S} from 1922 to 1991.  The plates were taken by astronomers at Maria Mitchell Observatory for a study of DF Cygni \citep{1984JAVSO..13...62B}, an RV Tau variable and the field of view of the plates fortunately encompasses KIC 8462852. KIC 8462852 is located $\sim$4.5$\degr$ from DF Cygni.  Throughout the period multiple plates were taken during single nights for several nights,  providing the opportunity to search for sudden dimming (dips) and brightening (flare) events that last only several days. We have extracted the photographic magnitudes of KIC 8462852 from 835 MMO plates dating from 1922 to 1991. The light curve of KIC 8462852 is used to search for such events, as well as the reported  long--term dimming.

\begin{figure}
	\centering
	\includegraphics[width=0.8\columnwidth]{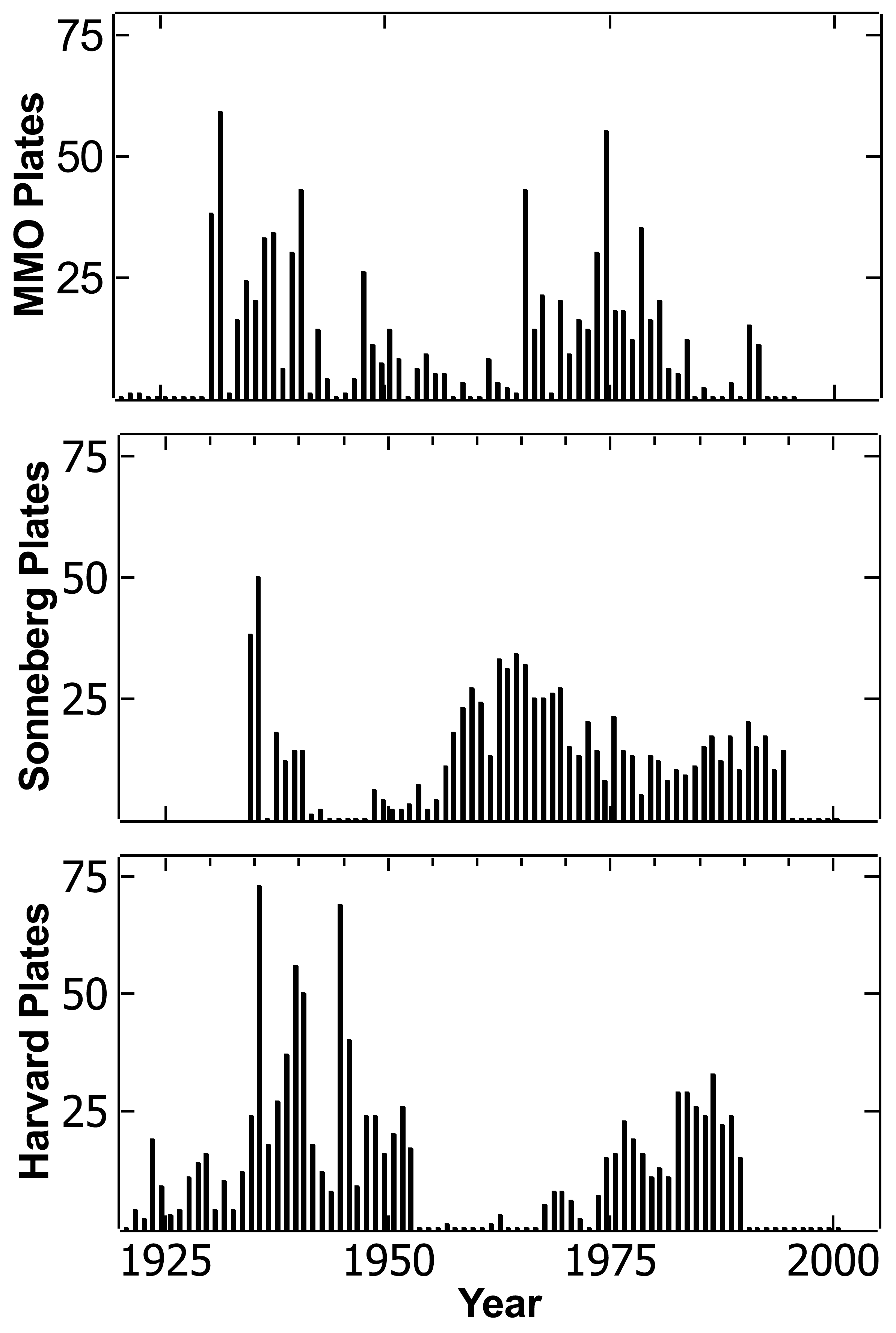}
    \caption{Histogram showing the number of plates that contain KIC 8462852 taken per year for the MMO, Harvard, and Sonneberg collections. }
    \label{fig:figure_1}
\end{figure}

\section{Data}

Figure~\ref{fig:figure_1} shows the number of MMO plates that contain KIC 8462852 taken per year from 1920 to 2000.  Also shown are the histograms for the Harvard plates \citep{2016ApJ...822L..34S} and Sonneberg plates \citep{2016ApJ...825...73H} that also contain KIC 8462852.  The coverage of the MMO plates is more heavily weighted towards the 1930's whereas the Sonneberg collection is more heavily weighted towards the 1960's. The MMO, Harvard, and Sonneberg collections complement each other.  Harvard plates before 1920 are not included in the histogram because we are interested in the comparison of the MMO plates with the other collections.  Note the MMO and Sonneberg collections fill in the Harvard "Menzel gap" from 1952--1965.

\begin{figure}
	\centering
	\includegraphics[width=0.8\columnwidth]{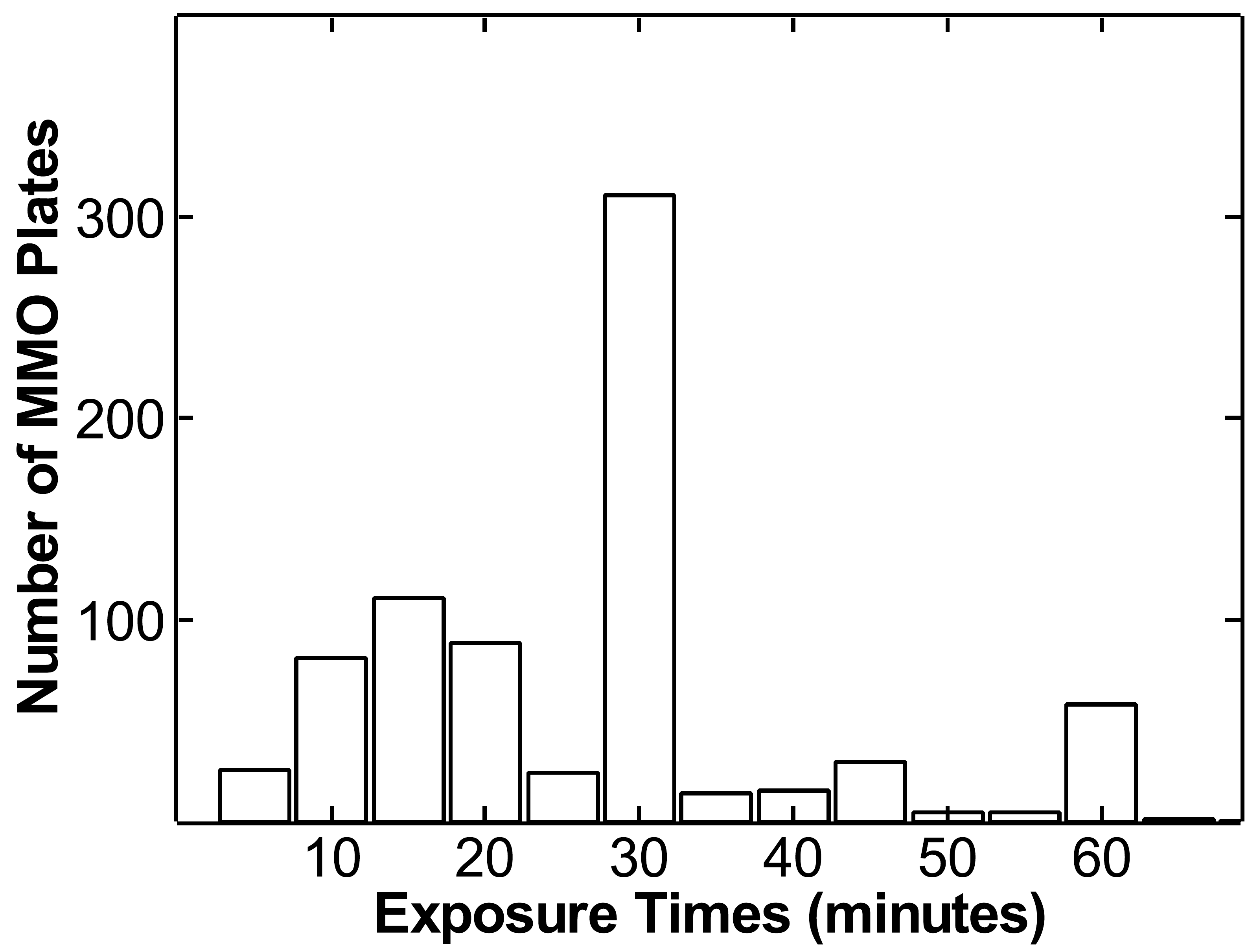}
    \caption{Histogram of exposure times of the plates used in this study. The two MMO plates taken in 1922 with exposure times of 180 minutes are not plotted.}
    \label{fig:figure_2}
\end{figure}

\subsection{The Maria Mitchell Plate Collection}
The Maria Mitchell Observatory \citep{2009ASPC..410...96S} photographic plate collection began in 1913 with the installation of the 7.5--inch Cooke/Clarke telescope. Plates are 8 x 10 inches with a plate scale of 240$\arcsec/mm$ and a field size of 13.5$\degr$ x 17$\degr$ providing a uniform set of images. A large majority of plates are blue sensitive \citep{1992JAVSO..21...99f}.  Exposure times for the plates in this study vary from 5 minutes to 60 minutes, with 2 plates (plate index numbers NA 468 and NA 487) with exposure times of 180 minutes taken in 1922 (Figure~\ref{fig:figure_2}). 

\begin{figure}
	\centering
    \includegraphics[width=0.9\columnwidth]{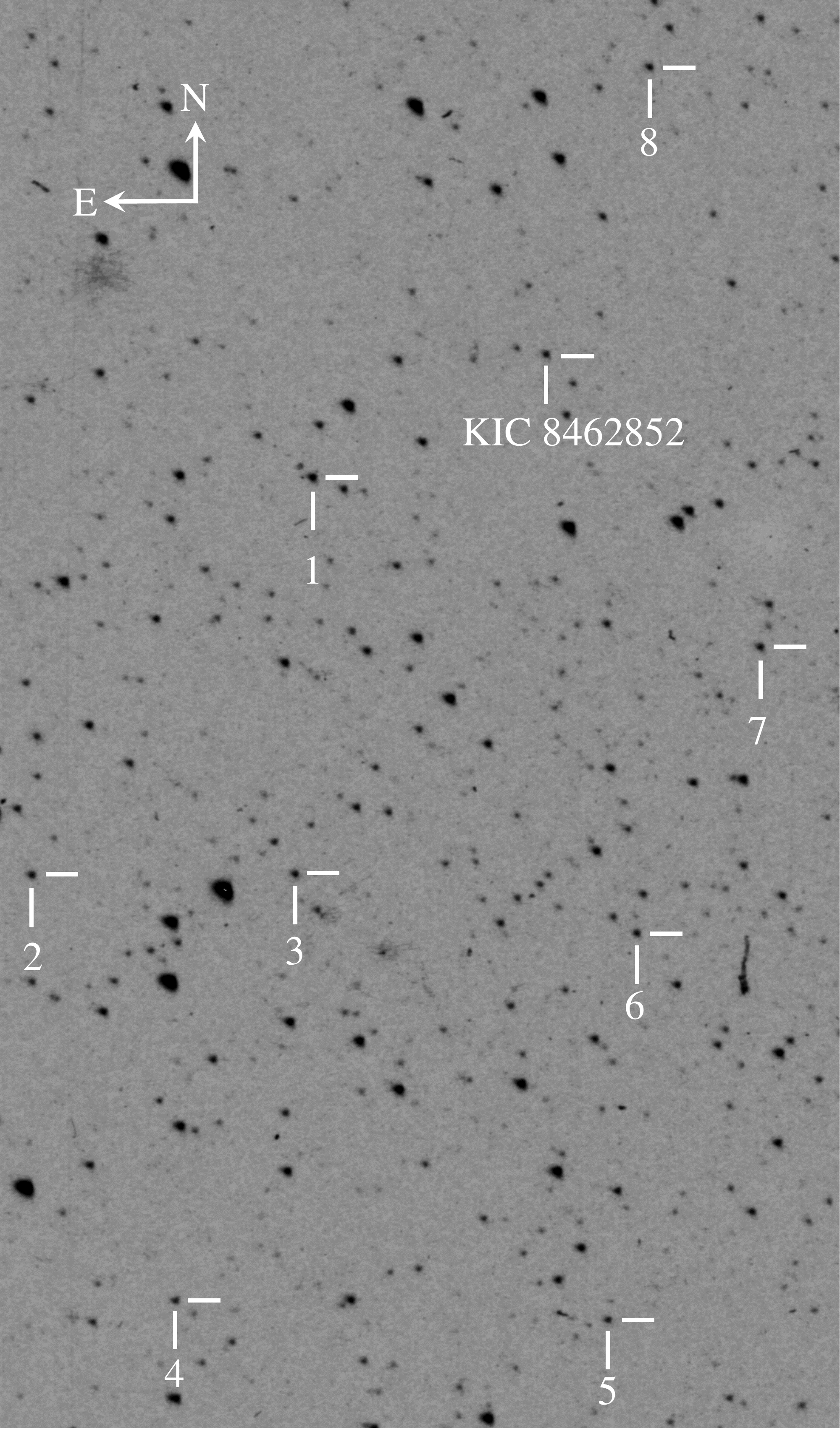}
    \caption{Finding chart. This field of view is from plate NA1143 taken 25 February 1931. The digitized image size is 10 mm x 17 mm (40$\arcmin$ x 68$\arcmin$).  KIC 8462852 and comparison stars are indicated in the figure and the comparison star names, coordinates and APASS magnitudes are given in Table~\ref{tab:table_1}.}
    \label{fig:figure_3}
\end{figure}

\subsection{KIC 8462852 Photometry}
\label{sec:KIC8462852-data} 

A 40$\arcmin$ x 68$\arcmin$ area around KIC 8462852 was digitized and 8 comparison stars in that area were selected. The criteria for comparison star selection was adopted from \cite{2016ApJ...822L..34S}, although the stars themselves are not necessarily the same comparison stars used in the Harvard study because of the much smaller field of view used for the MMO photometry.  Comparison stars are within one spectral subclass, and within 0.5 visual magnitudes of KIC 8462852.  The comparison stars are also selected because they are not identified as variables after a search through literature.  Eight stars meet these criteria. The comparison stars are listed in Table~\ref{tab:table_1}. Figure~\ref{fig:figure_3} is the finding chart for the KIC 8462852 field of comparison stars. In this paper, to be consistent with the Harvard photometry of KIC 8462852 \citep{2016ApJ...822L..34S}, the comparison star photometric magnitudes are taken from
the AAVSO Photometric All-Sky Survey \citep[APASS;][]{2014CoSka..43..518H}. Prior to 1950, three emulsion types were used that included Speedway, Cramer Presto, and Cramer Hispeed and after 1950, only Eastman Kodak 103aO and IIaO emulsions were used \citep{2004JAVSO..32..126D}.  All of these emulsions are blue sensitive, and the APASS magnitudes are closest in bandpass to the emulsions. However, because the MMO plates do have a variety of emulsion types, we refer to the MMO photometric results as photographic magnitudes ($m_{pg}$).

\begin{deluxetable}{ccccc} 
	\tablecaption{Comparison Stars.\tablenotemark{a}\label{tab:table_1}}
	\tablehead{
    \colhead{Star} & \colhead{TYC} & 
    \colhead{RA} & \colhead{Dec} & \colhead{APASS} \\
    \colhead{No.} &\colhead{3162-} & 
    \colhead{(J2000)} & \colhead{(J2000)} & \colhead{Magnitude}  
    }
	\startdata
	1 & 1320-1 & 20h 07m 09.068s & 44$\degr$ 20$\arcmin$ 17.06$\arcsec$ & 12.133\\
	2 & 1698-1 & 20h 08m 28.312s & 44$\degr$ 00$\arcmin$ 36.23$\arcsec$ & 12.108\\
        3 & 488-1 & 20h 07m 16.759s & 44$\degr$ 01$\arcmin$ 24.86$\arcsec$ & 12.158\\
        4 & 964-1 & 20h 07m 42.879s & 43$\degr$ 40$\arcmin$ 14.10$\arcsec$ & 12.478\\
        5 & 420-1 & 20h 05m 45.801s & 43$\degr$ 40$\arcmin$ 28.10$\arcsec$ & 12.604\\
        6 & 462-1 & 20h 05m 43.265s & 43$\degr$ 59$\arcmin$ 27.68$\arcsec$ & 12.351\\
        7 & 316-1 & 20h 05m 13.017s & 44$\degr$ 13$\arcmin$ 41.71$\arcsec$ & 12.125\\
        8 & 509-1 & 20h 05m 50.997s & 44$\degr$ 41$\arcmin$ 43.34$\arcsec$ & 12.107\\
    \enddata
    \tablenotetext{a}{Figure 3 shows the field of view with these comparison stars.}
\end{deluxetable}

The photometry follows the method used by the Harvard DASCH project  \citep{2009ASPC..410..101G,2013PASP..125..857T}.  Stars are extracted using the Source Extractor routine in the MIRA Pro photometry software  \citep{2007Icar..187..285W}.  We set the threshold to 10 sigma above the background and measure the effective area of the eight selected comparison stars. This approach is more similar to iris photometry, where stellar magnitudes are measured by a density weighted image diameter, than today's more common aperture photometry using total flux through a fixed measurement aperture. Every digitized image was visually inspected for defects like scratches on the emulsion.  A total of 867 plates were found in the MMO collection with images of KIC 8462852. Of these, 32 were rejected because of defects or faint stellar images due to exposure times less than 5 minutes. A linear fit between the comparison star  magnitudes and effective areas provides the calibration for KIC 8462852.  This method deviates from the DASCH method where all stars in the field from the brightest to the faintest are used for calibration which requires a more robust fitting algorithm between magnitude and effective area.  Because all of our comparison stars are within 0.5 mag of KIC 8462852, we used a linear fit.  

\begin{figure}[htb!]
	\centering
	\includegraphics[width=\columnwidth]{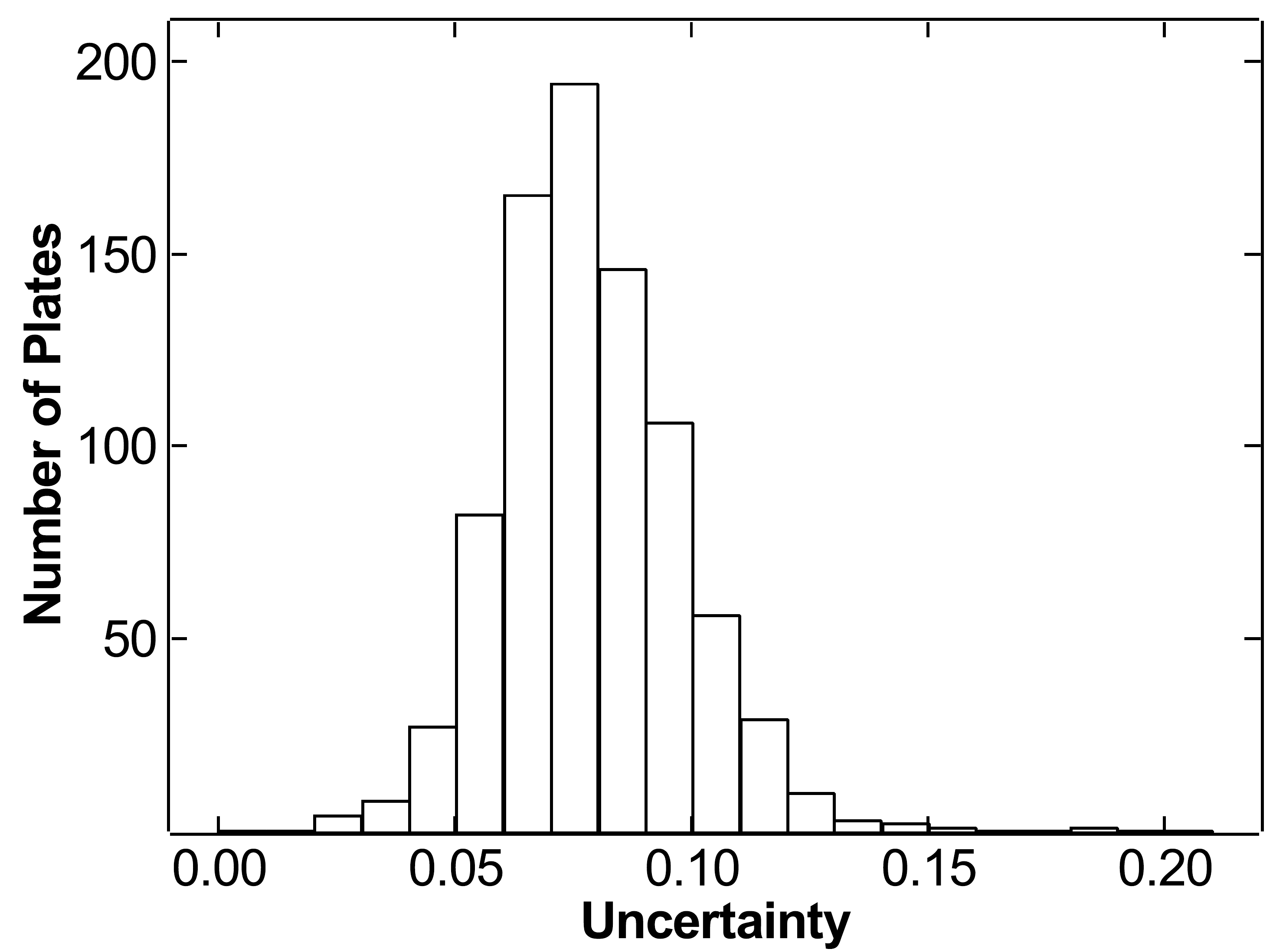}
    \caption{Histogram of uncertainties of the MMO Photometry.}
    \label{fig:figure_4}
\end{figure}

\begin{figure*}
    \includegraphics[width=\textwidth]{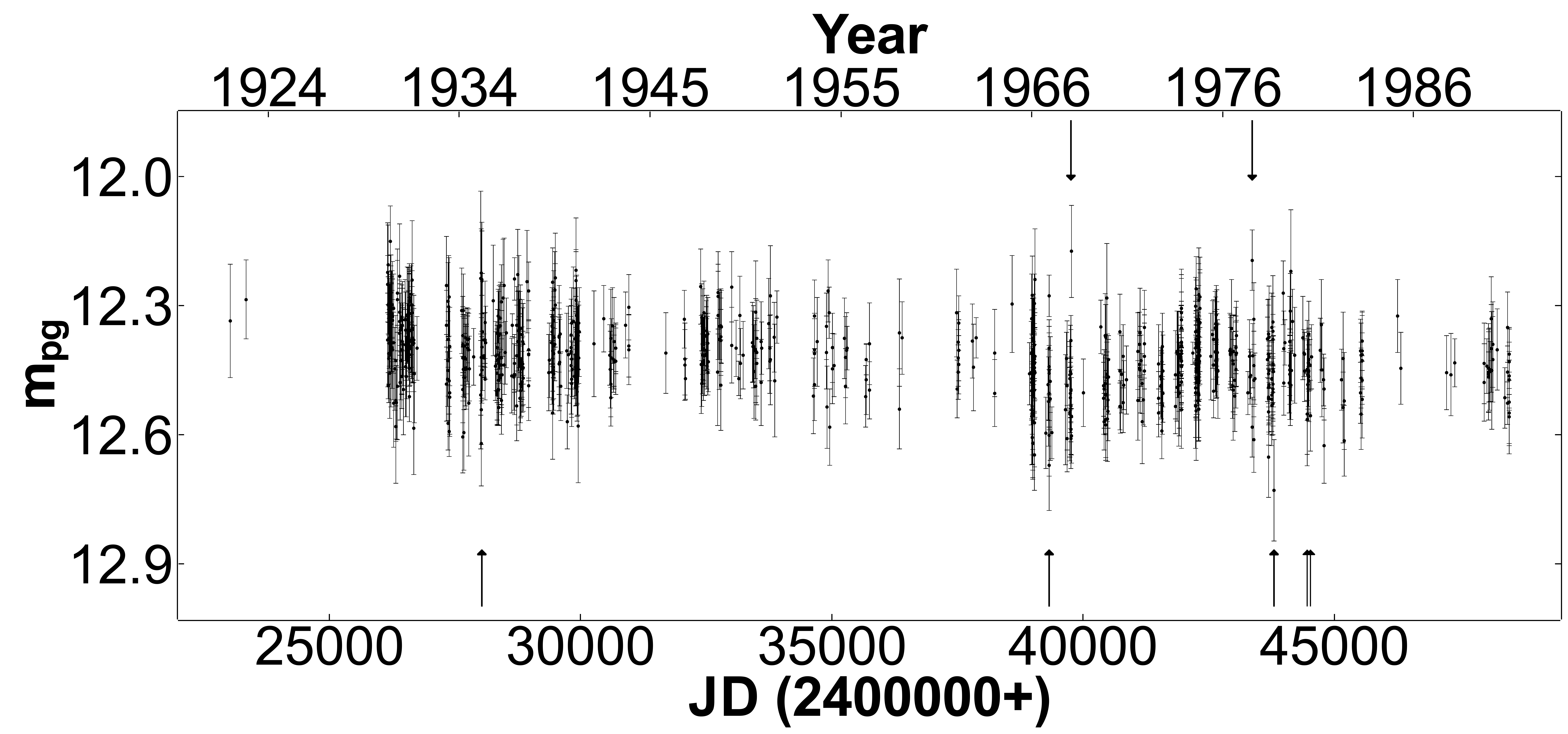}
    \caption{Light curve of KIC 8462852 from MMO photographic plates from 1922 to 1991. The up arrows point to dip events and the down arrows point to flare events. Two up arrows are blended for dip events which occurred between JD2444464 and 2444522 in 1980. For details see  section~\ref{sec:section3.2}, and Table~\ref{tab:table_3}}.
    \label{fig:figure_5}
\end{figure*}

From the linear fit of magnitude versus effective area we derive the magnitude of KIC 8462852.  
The mean residual is a measure of the uncertainty of the magnitude of KIC 8462852.   Figure~\ref{fig:figure_4} is a histogram of uncertainties of all measured plates.  The mean uncertainty is 0.07 mag and 81$\%$ of the plates have uncertainties $\leq$ 0.1 mag.  Figure~\ref{fig:figure_5} shows the resulting MMO light curve of KIC 8462852. 

For reproducibility and independent re-analysis, we
release the data used to produce the figures.\footnote{https://github.com/castelaz/kic8462852-MMO-data}

\begin{figure*}
	\includegraphics[width=\textwidth]{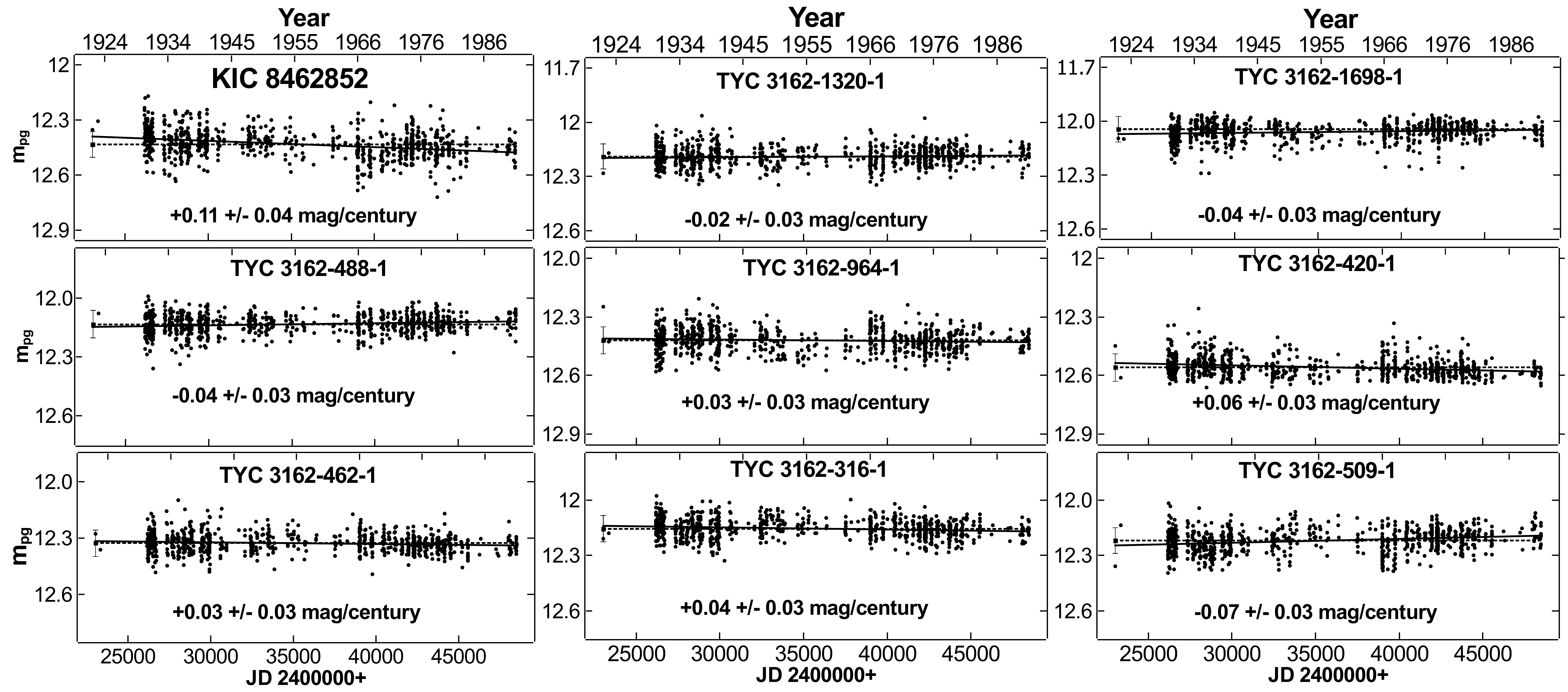}
    \caption{The light curve of KIC 8462852 and the comparison stars with linear fits to each light curve (solid lines). The slopes are given for each linear fit. A dotted line is drawn horizontally across each plot using the mean magnitude of each star, and a typical error bar is shown  for reference on the left of each light curve.}
    \label{fig:figure_6}
\end{figure*} 

\section{Results}

\subsection{Long Term Variation}
\begin{deluxetable}{ll}
	\tablecaption{Slopes of light curves of KIC 8462852 and comparison stars.\label{tab:table_2}}
	\tablehead{
	\colhead{Star} & \colhead{Slope} \\
    \colhead{} & \colhead{(mag/century)} 
	}
	\startdata
		KIC 8462852 & $+0.11\pm0.04$ \\
		TYC 3162-1320-1 & $-0.02\pm0.03$ \\
        TYC 3162-1698-1 & $-0.04\pm0.03$ \\
        TYC 3162-488-1 & $-0.04\pm0.03$ \\
		TYC 3162-964-1 & $+0.03\pm0.03$ \\
        TYC 3162-420-1 & $+0.06\pm0.03$ \\
        TYC 3162-462-1 & $+0.03\pm0.03$ \\
		TYC 3162-316-1 & $+0.04\pm0.03$ \\
        TYC 3162-509-1 & $-0.07\pm0.03$ \\
	\enddata
\end{deluxetable}

For perspective, Figure~\ref{fig:figure_6} shows the light curves of KIC  8462852 and the comparison stars.  A linear fit to each light curve is included in Figure~\ref{fig:figure_6}. Table~\ref{tab:table_2} shows the slopes from the linear fits for KIC 8462852 and the comparison stars.  The absolute values of the slopes of 6 of the 8 comparison stars are $\leq$0.04$\pm$0.03 mag/century. One comparison star has a slope of +0.06$\pm$0.03, and another has a slope of -0.07$\pm$0.03 mag/century. In contrast, the derived slope of KIC 8462852 is +0.11$\pm$0.04 mag/century, larger than any of the comparison stars.

\begin{deluxetable*}{cccccc}
	\tablecaption{Dip and Flare events detected in the MMO light curve (Figure~\ref{fig:figure_5}). Because the 2013 Kepler data shows dips separated by up to 50 days, the nearest minima that appear in the Harvard, Sonneberg, and Sternberg data are given when the extrema are within $\pm$50 days of the MMO extrema.  The Sonneberg data are differential magnitudes in the two color bands, pv (red) and pg (blue) \citep{2016ApJ...825...73H}. Blanks are given where photographic data does not exist.\label{tab:table_3}}
	\tablehead{
    \colhead{} & \colhead{MMO} & \colhead{Harvard} &
    \colhead{Sonneberg} & \colhead{Sonneberg} & \colhead{Sternberg} \\
    \colhead{} & \colhead{} & \colhead{} & \colhead{pv} & \colhead{pg} & \colhead{} 
    }
	\startdata
	Average Mag & 12.43 & 12.34  & -0.01 & 0.00 & 12.26 \\
	& & & &  &  \\
	Dips & & & &  &   \\
\tableline
		JD & 2428036 & 2428017 & \nodata & 2428054 & \nodata \\
        Date & 21 Aug 1935 & 02 Aug 1935 & \nodata & 8 Sep 1935  & \nodata \\
        Mag	& 12.61$\pm$0.11 & 12.87$\pm$0.26 & \nodata & 0.39$\pm$0.23  & \nodata  \\
\tableline 
		JD & 2439323 & \nodata & 2439363 &  \nodata   &  \nodata \\
        Date & 16 Jul 1966 & \nodata & 25 Aug 1966   & \nodata &   \nodata\\
        Mag	& 12.66$\pm$0.09 & \nodata & 0.49$\pm$0.13  & \nodata    & \nodata \\
\tableline
        JD & 2443803 & 2443792   & \nodata  & \nodata & 2443806 \\
        Date & 21 Oct 1978 & 10 Oct 1978   & \nodata  & \nodata &  24 Oct 1978 \\
        Mag & 12.68$\pm$0.11 & 12.69$\pm$0.21  & \nodata & \nodata &  12.36$\pm$0.10 \\
\tableline
        JD & 2444464 & 2444467  & \nodata &  \nodata  & \nodata \\
        Date & 12 Aug 1980 & 15 Aug 1980  &  \nodata   & \nodata   &  \nodata\\
        Mag & 12.69$\pm$0.07 & 12.45$\pm$0.14  &  \nodata  &  \nodata   & \nodata\\
        & & &  & & \\
\tableline
        JD & 2444522 & 2444523  & 2444521 & 2444521 &  \nodata\\
        Date & 09 Oct 1980 & 10 Oct 1980  & 08 Oct 1980 & 8 Oct 1980  & \nodata  \\
        Mag & 12.56$\pm$0.07 & 12.61$\pm$0.12 & 0.25$\pm$0.19 & 0.42$\pm$0.15    & \nodata\\  
        &  & &  & & \\
	Flares & & &   & & \\	
\tableline
		JD & 2439764 & 2439733  & 2439735  & 2439735 & \nodata  \\
        Date & 30 Sep 1967 & 30 Aug 1967  & 01 Sep 1967  & 01 Sep 1967 & \nodata \\
        Mag	& 12.20$\pm$0.09 & 12.29$\pm$0.11  & -0.29$\pm$0.16  & -0.13$\pm$0.11 & \nodata  \\
\tableline
       JD & 2443366 & 2443371  & 2443372 & 2443372 & \nodata \\
       Date & 10 Aug 1977 & 15 Aug 1977  & 16 Aug 1977  & 16 Aug 1977 & \nodata \\
        Mag & 12.22$\pm$0.06 & 12.20$\pm$0.11  & -0.48$\pm$0.13 & -0.10$\pm$0.10 & \nodata \\      
    \enddata
\end{deluxetable*}

\begin{figure}
	\centering
	\includegraphics[width=0.7\columnwidth]{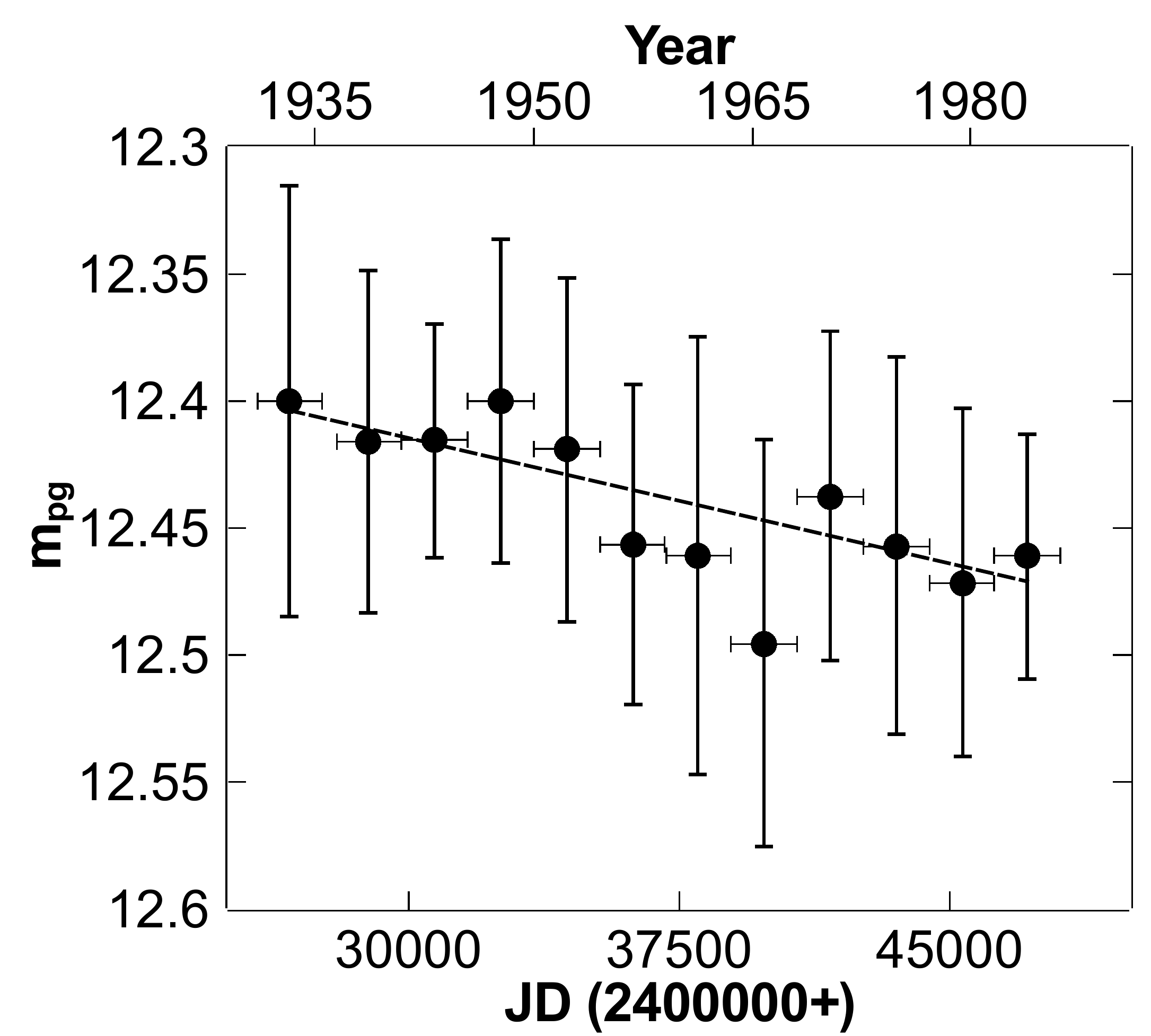}
    \caption{The light curve of KIC 8462852 binned in 5 year intervals.  The slope of the light curve is 0.12 $\pm$0.02 magnitudes/century (dashed line).  The x-axis error bars show the 5 year span for each data point. The y-axis error bars are the average residuals to the fit.}
    \label{fig:figure_7}
\end{figure}

The MMO light curve (Figures~\ref{fig:figure_5} and \ref{fig:figure_6}) indicate a decrease in magnitude trend of +0.11 $\pm$ 0.04 mag/century.  Although not as steep as the 0.164$\pm$0.013 magnitude/century slope measured by \cite{2016ApJ...822L..34S}, the slope measured from the 70 year MMO light curve is consistent with the Harvard data.  \cite{2016ApJ...822L..34S} had binned the Harvard data in 5 year intervals and doing the same with the MMO data, the linear fit indicates a decrease of +0.12$\pm$0.02 mag/century  (Figure~\ref{fig:figure_7}). The decrease in magnitude from 1922 to 1991 shown in Figure~\ref{fig:figure_7} clearly indicates that KIC 8462852 is slowly fading.  These results contrast with the analysis of \cite{2016ApJ...825...73H} who cite a slope of 0.046$\pm$0.040 mag/century from all of the available Sternberg data and 0.09$\pm$0.02 from Sternberg Astrograph data alone.  However, the Astrograph data is not well-sampled, and only 17 measurements exist between 1960 and 1990.   The MMO light curve supports the observation of slow fading of KIC 8462852.

\subsection{Episodes of Dip and Flare Events}\label{sec:section3.2}

Several photometric data points in the light curve of KIC 8462852 appear to stand out either as much dimmer (dips) or much brighter (flares) than most data points in the light curve.  These anomalous data points are designated in Figure~\ref{fig:figure_5} with arrows.  Table~\ref{tab:table_3} lists the Julian Dates, dates, and magnitudes of these apparent dip and flare events in the MMO light curve. The photographic magnitudes of these events are fainter than 12.6 and brighter than 12.2, $\sim$15$\%$ different than the average magnitude of 12.4, or 2 to 3 sigma difference based on the range of uncertainties of 0.06 magnitudes to 0.11 magnitudes of the measurements. 

\begin{figure}
	\centering
	\includegraphics[width=0.9\columnwidth]{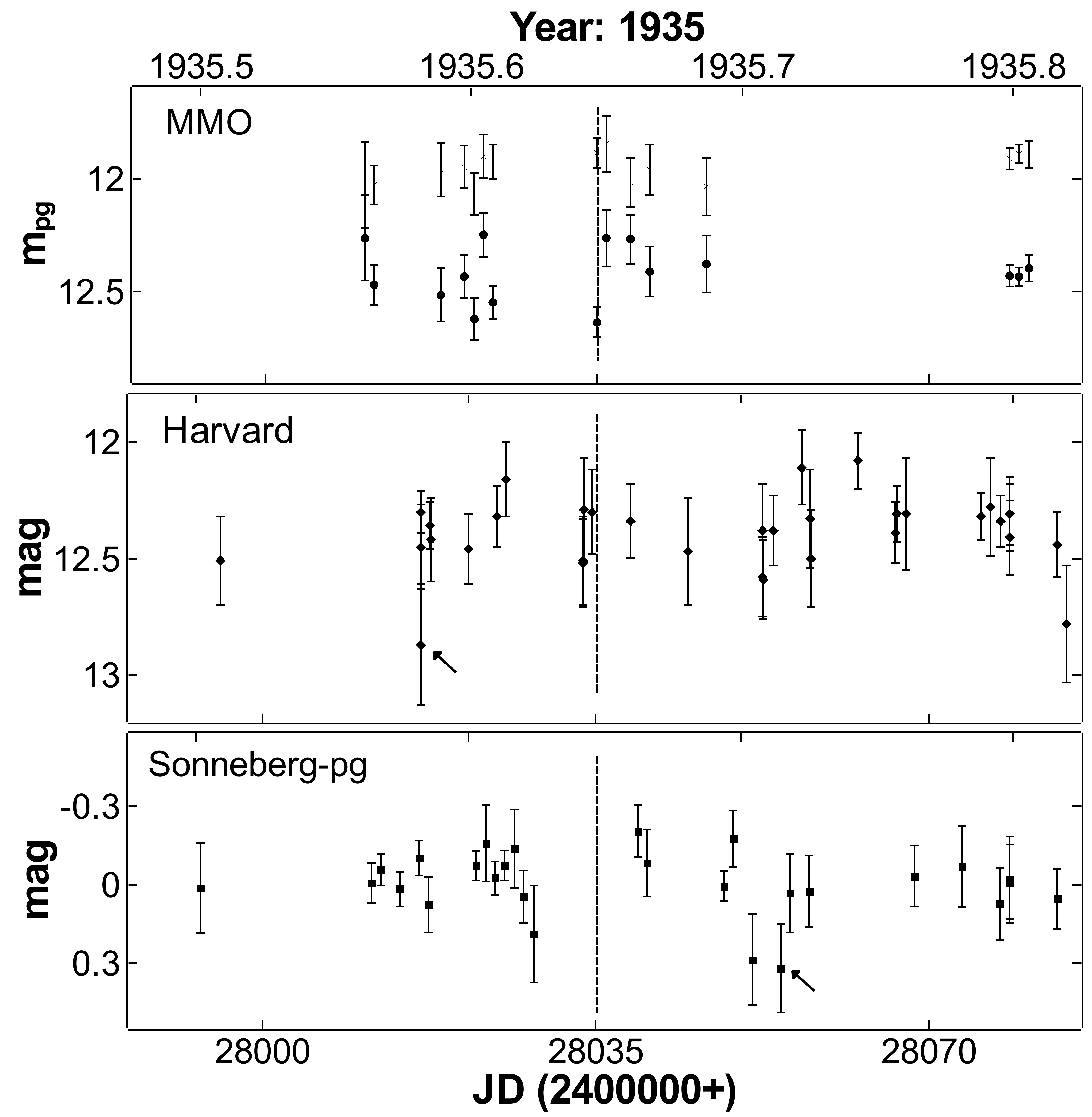}
    \caption{MMO light curve showing a dip event in 1935. The time spans 50 days before the MMO dip to 50 days after the MMO dip. The median comparison star magnitudes (x's), offset by m$_{pg}$ = -0.25, don't show a dip. The MMO dip occurs on 21 August 1935 (1935.638, JD 2428036). Harvard and Sonneberg-pg data exists and those data are also shown.  A dip in the Harvard data appears to occur on 2 August 1935 (arrow), and 3 dips appear to occur in the Sonneberg data with the deepest on 8 September 1935 (arrow). Sonneberg-pv data does not exist in 1935. The dotted vertical line marks the MMO dip event in each plot. See Table~\ref{tab:table_3} for dates and magnitudes.}
    \label{fig:figure_8}
\end{figure}

\begin{figure}
	\centering
	\includegraphics[width=0.9\columnwidth]{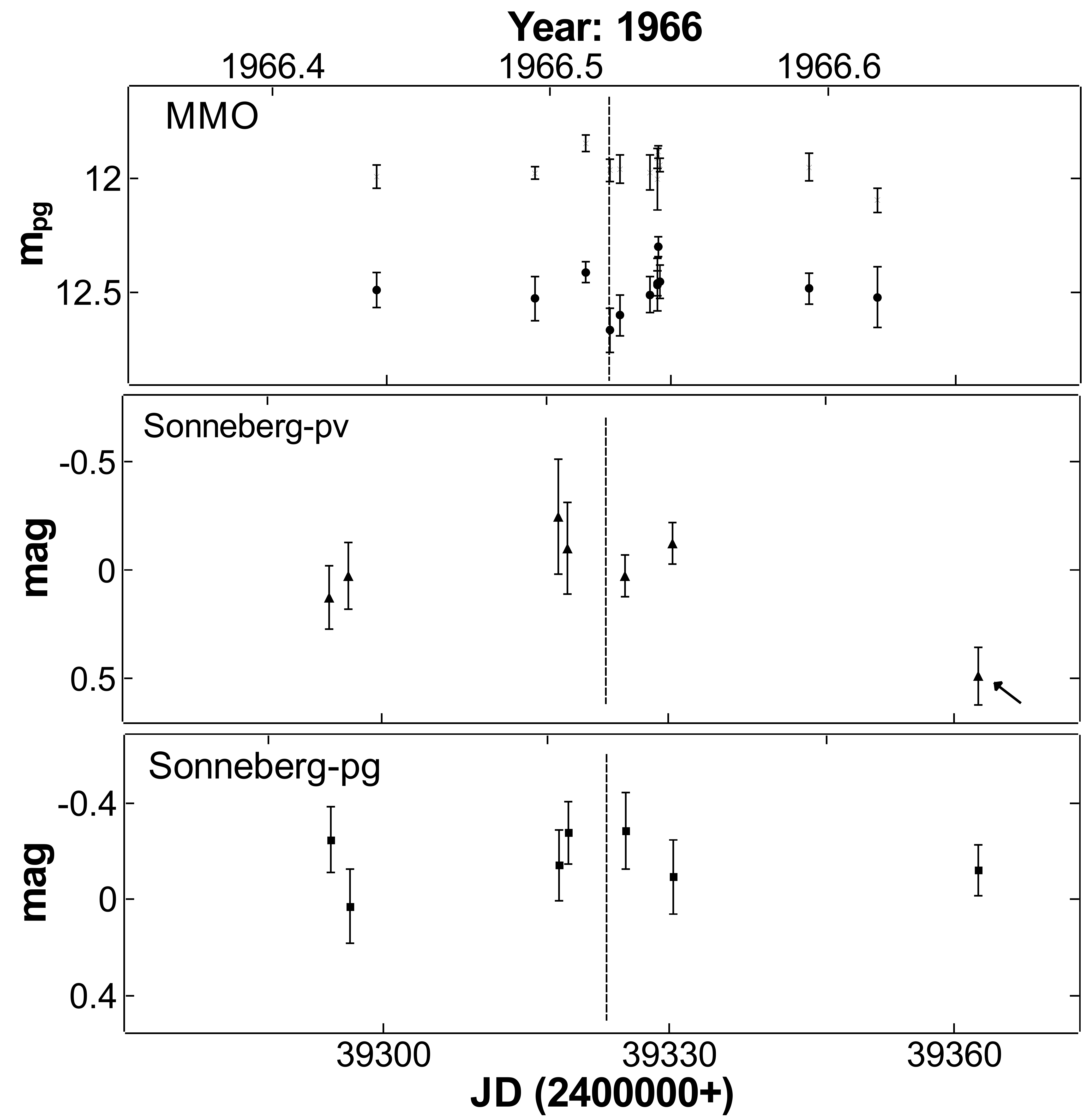}
    \caption{MMO light curve showing a dip event in 1966. The time spans 50 days before the MMO dip to 50 days after the MMO dip event. The median comparison star magnitudes (x's), offset by m$_{pg}$ = -0.25, don't show a dip. The MMO dip occurs on 16 July 1966 (1966.512, JD 2439323). Sonneberg-pv and Sonneberg-pg data exists and those data are also shown.  A dip event appears to occur in the Sonneberg-pv data on 25 August 1966 data (arrow), but the Sonneberg-pg data does not show a dip event on or around 25 August 1966. The dotted vertical line marks the MMO dip event in each plot. See Table~\ref{tab:table_3} for dates and magnitudes.}
    \label{fig:figure_9}
\end{figure}

\begin{figure}
	\centering
	\includegraphics[width=0.9\columnwidth]{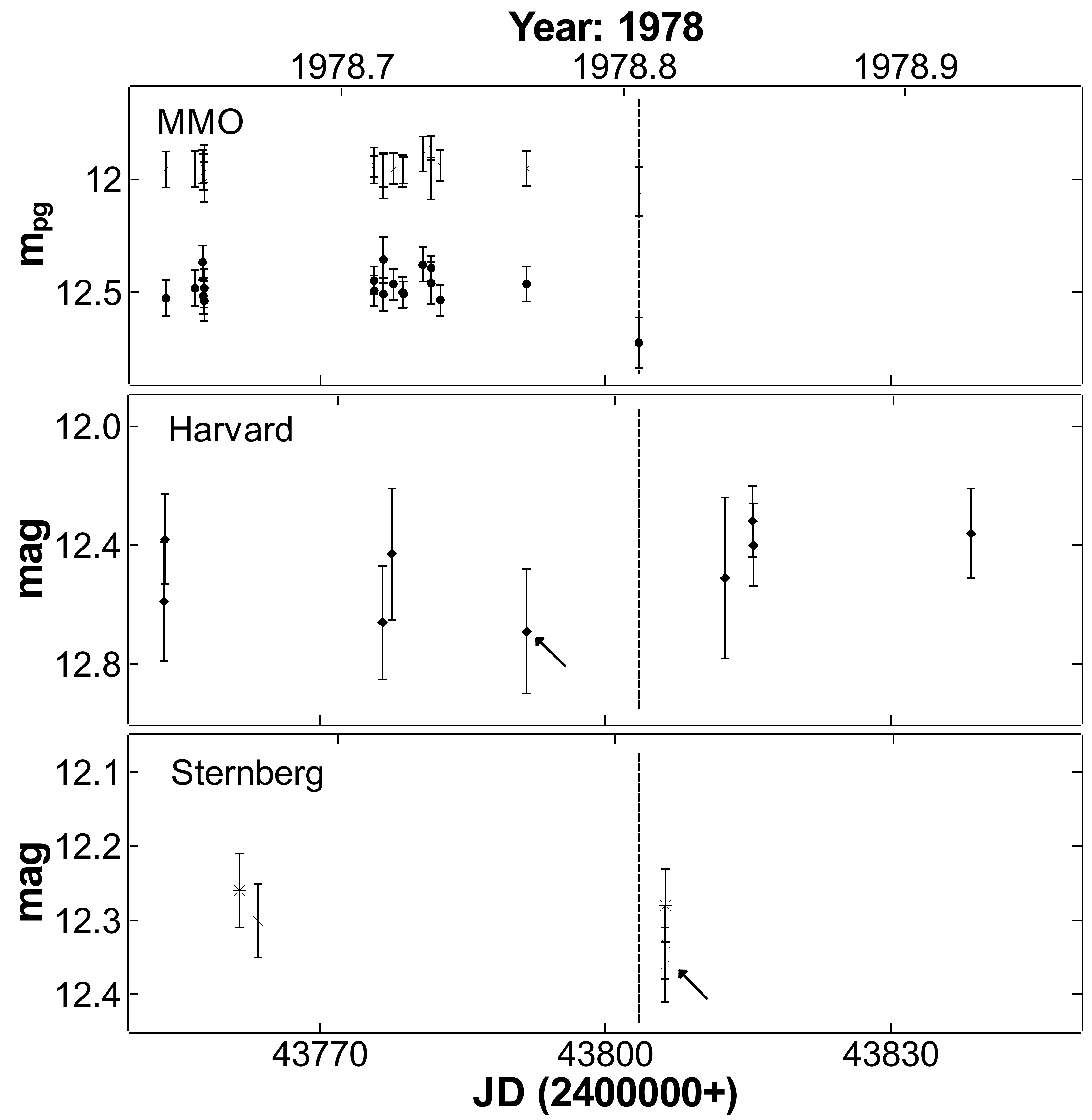}
    \caption{MMO light curve showing a dip event in 1978. The time spans 50 days before the MMO dip to 50 days after the MMO dip event. The median comparison star magnitudes (x's), offset by m$_{pg}$ = -0.25, don't show a dip. The MMO dip occurs on 21 October 1978 (1978.805, JD 2443803). Harvard and Sternberg data exists and those data are also shown.  A dip  appears to occur in the Harvard data a few days before on 10 October 1978 (arrow), and a few days after in the Sternberg data on 24 October 1978 (arrow). Sonneberg-pv and Sonneberg-pg data don't exist for this time period. The dotted vertical line marks the MMO dip event in each plot. See Table~\ref{tab:table_3} for dates and magnitudes.}
    \label{fig:figure_10}
\end{figure}

\begin{figure}
	\centering
	\includegraphics[width=0.9\columnwidth]{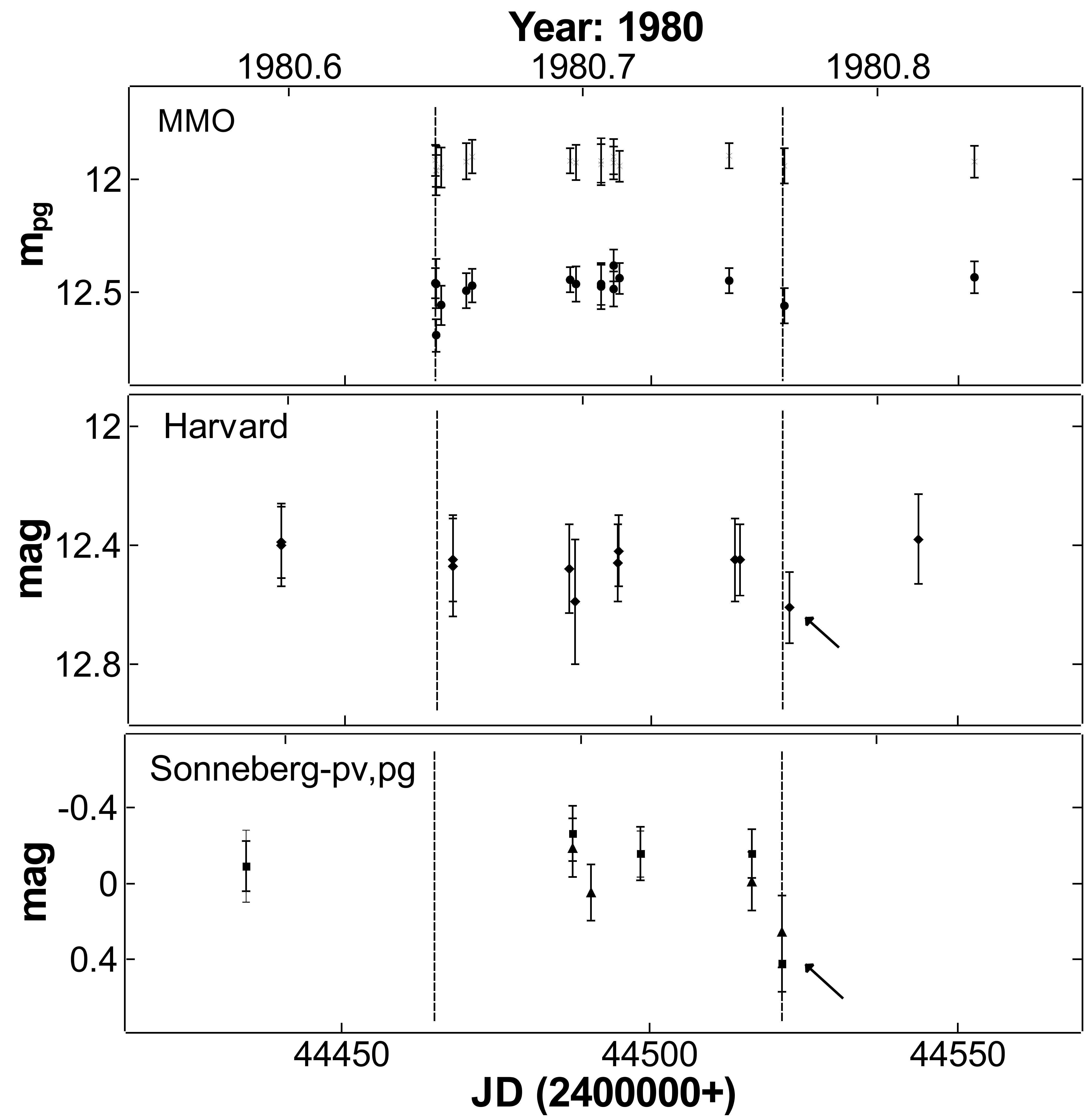}
    \caption{MMO light curve showing two dip events in 1980. The time spans 50 days before the first MMO dip to 50 days after the second MMO dip event. The median comparison star magnitudes (x's), offset by m$_{pg}$ = -0.25, don't show a dip. The MMO dips occurs on 13 August 1980 (1980.617, JD 2444465) and 9 October 1980 (1980.773, JD 2444522). Harvard, Sonneberg-pv (triangles), and Sonneberg-pg (rectangles) are also shown.  An apparent dip in the Harvard data on 10 October 1980 (arrow), occurs one day after the second MMO dip event. The Sonneberg-pv and Sonneberg-pg data only exist in October and dips appear on 8 October 1980 (arrow). The dotted vertical lines marks the MMO dip events in each plot. See Table~\ref{tab:table_3} for dates and magnitudes.}
    \label{fig:figure_11}
\end{figure}

\begin{figure*}
	\centering
	\includegraphics[width=\textwidth]{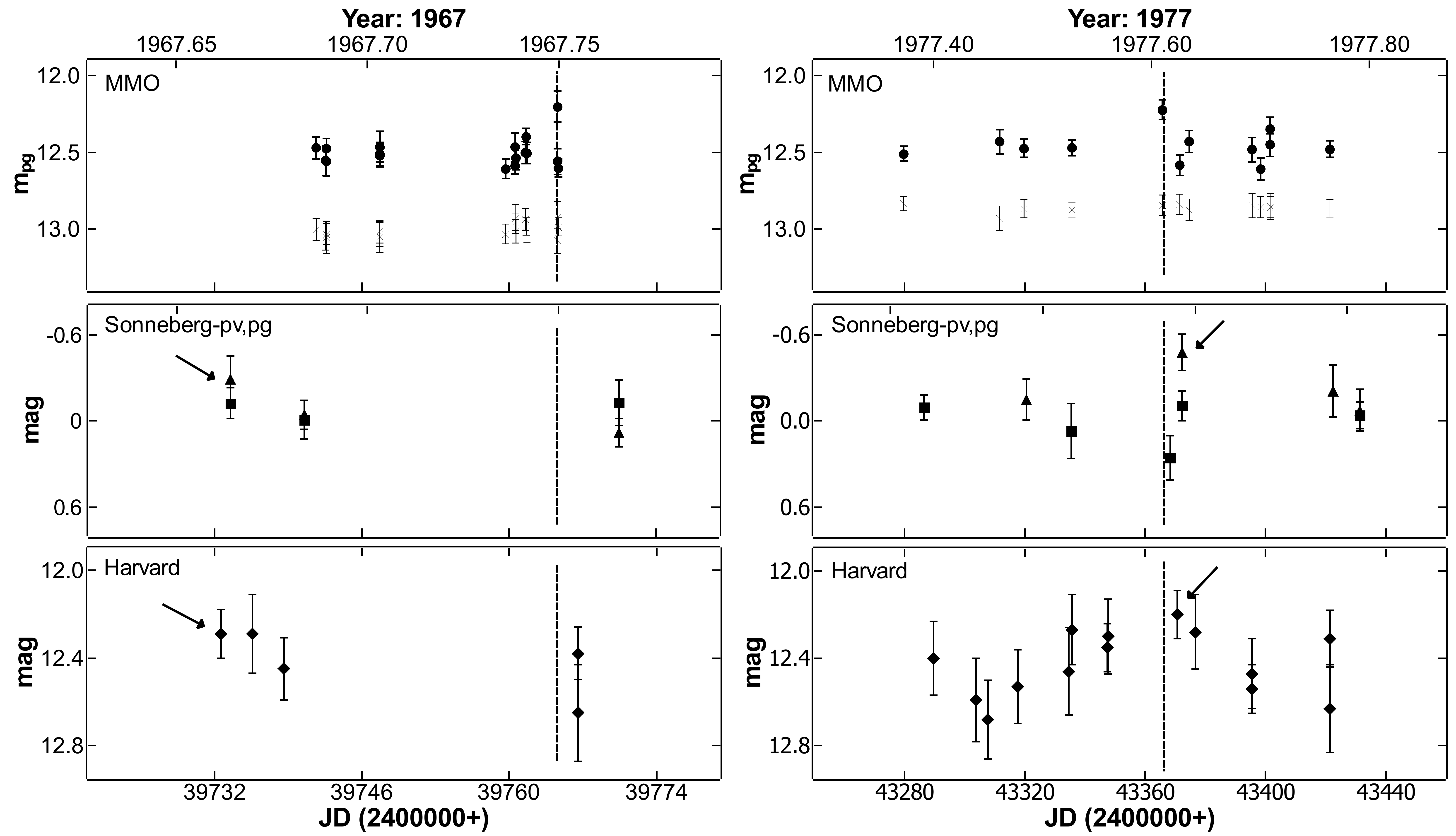}
    \caption{MMO, Harvard, Sonneberg-pv (triangles), and Sonneberg-pg (rectangles) light curves of observed flares over the period 30 August - 30 September 1967, and again 10-16 August 1977. The dotted vertical line marks the MMO flare events in each plot. Flares observed in the Harvard and Sonneberg light curves are indicated by arrows. Sternberg data doesn't exist for these dates. The median MMO comparison star magnitudes (x's; top frame), with magnitudes offset by  m$_{pg} = +0.7$, do not show the increase seen in KIC 8462852. See Table~\ref{tab:table_3} for dates and magnitudes.}
    \label{fig:figure_12}
\end{figure*}

If the MMO events listed in Table~\ref{tab:table_3} are real, then we would expect light curves from other data sets (Harvard, Sonneberg, and Sternberg) to show dip and flare events during the same time periods.  Because the Kepler 2013 light curve shows a number of dips over a 50 day period, we also search a time window of $\pm$50 days around each MMO event for dips and flares in the other data sets (Harvard and Sonneberg plate collections). Also, keeping in mind that the observations conducted in each data set were not synchronized with each other, and in most cases data may have been taken for several nights, and then gaps of days or weeks before the next set, we would not expect to find dip or flare events between data sets to exactly match.  In fact, the dip and flare events may not actually be the faintest or brightest variation.  Only regular, multiple observations over a long time frame, like the Kepler data, will measure the true dip or flare event magnitude.  With these caveats in mind, comparison to Harvard \citep{2016ApJ...822L..34S}, Sonneberg and Sternberg   \citep{2016ApJ...825...73H} light curves with dip events within $\pm$50 days of the corresponding MMO events are given in Table~\ref{tab:table_3}.  The Harvard, Sonneberg, and Sternberg magnitudes are each 0.2 magnitude fainter or brighter than their respective average magnitudes, and represent $\sim$15$\%$ differences.  Figures~\ref{fig:figure_8}, \ref{fig:figure_9}, \ref{fig:figure_10}, and \ref{fig:figure_11} show the dip events detected in 1935, 1966, 1978, and 1980, respectively.

The observed dip and flare events could be due to factors related to the night sky, image quality, and exposure time, for example, and should not be dismissed. However, visual inspection of the MMO plates does not show defects or dirt near KIC 8462852 or any of the comparison stars. Also, the effects of sky conditions and image quality on the photometry is minimized because the 8 comparison stars are near KIC 8462852 and would be affected in the same way. So, the dip and flare light curves shown in Figures 8-12 and given in Table~\ref{tab:table_3} are taken as real.

Figure~\ref{fig:figure_8} shows the MMO light curve for a dip event that occurs on 21 August 1935 (Table~\ref{tab:table_3}). Data on the Harvard light curve was taken two days before the MMO dip event and does not show a  dip.  However, a dip event appears to occur in the Harvard data on 2 August 1935. The Sonneberg-pg data shows 3 possible dip events near the MMO dip event, two occurring on 8 September  1935 with relative magnitudes of 0.39$\pm$0.23 and 0.31$\pm$0.16. The third possible dip event in the Sonneberg-pg data occurs at 0.19$\pm$0.19.  Because of the irregular spacing in time, none of these events correlate with any others, but do occur within 50 days of each other, and may be similar to events observed in the Kepler 2013 light curve that shows multiple dip events.  

The 16 July 1966 dip event observed in the MMO light curve (Figure~\ref{fig:figure_9}; Table~\ref{tab:table_3}) is $m_{pg}$ = 12.66$\pm$0.09 and seems to continue with $m_{pg}$ = 12.61$\pm$0.09 on 17 July 1966.  But, by 20 July 1966, the MMO light curve no longer shows a dip.   The Sonneberg-pv and Sonneberg-pg light curves have data on 12 July 1966 and 19 July 1966, and neither of these data show dip events similar to the July dates in the MMO light curve where no dips occur. Note the dip in the Sonneberg-pv data on 25 August 1966, 39 days after the MMO dip. A similar dip event would be expected in the Sonneberg-pg data taken on the same date, but one is not observed.  MMO data does not exist for 25 August 1966.  So, the nature of the apparent Sonneberg-pv dip event on 25 August 1966 is unclear. 

The dip event of $\sim$10$\%$  shown in Figure~\ref{fig:figure_10} and given in Table~\ref{tab:table_3} on 22 October 1978 occurs in the last set of plates (plate index numbers NA6114, NA6115, and NA6116) taken at the end of the 1978 season. Photometry for the next season beginning 23 April 1979 shows the magnitude is once again like that from earlier in 1978 (see Figure~\ref{fig:figure_5}).  The fact that the Sternberg data shows a drop on 24 October 1978 \citep{2016ApJ...825...73H}, and the Harvard data drops on 10 October 1978 suggests the event is real.  Sonneberg  data do not exist during this time period.  Being aware that the photographic plates were taken with a non-periodic cadence, with several plates taken one night followed by several nights of no imaging, the 1978 event may be near a minimum, but is not necessarily the minimum.  

We observe 2 dip events in 1980 separated by 57 days (Figure~\ref{fig:figure_11}; Table~\ref{tab:table_3}).  One occurs on 12 August 1980. Three MMO photographic plates were taken on 12 August 1980 (JD 2444464.69, JD 2444464.80, and JD 2444464.83).  Photometry from the first two plates, separated by 158 minutes, give $m_{pg}$ = 12.45$\pm$0.07 and $m_{pg}$ = 12.46$\pm$0.10, respectively. The third in the sequence for that night of observing, 45 minutes after the second, gives $m_{pg}$ = 12.69$\pm$0.07. The next observations were on 13 August 1980 and 17 August 1980 and these give $m_{pg}$ = 12.56$\pm$0.08 and $m_{pg}$ = 12.49$\pm$0.08, respectively, indicating a return to an average MMO magnitude.  The nearest Harvard data is mag = 12.45$\pm$0.21 on 15 August 1980, which does not appear to be significantly different from the Harvard average magnitude of 12.34, and is consistent with the 17 August 1980 MMO magnitude returning to an average magnitude. This suggests the 12 August 1980 dip event lasted only a few days. 

The second MMO dip in 1980 occurred on 9 October.  Note that Harvard, and Sonneberg light curves have data on 10 October 1980 and 8 October 1980, respectively, so a comparison in magnitudes can be made between the data sets separated by only 2 days.  The MMO dip is $\sim$5$\%$ below the average, whereas the Harvard and Sonneberg magnitudes are  $\sim$10$\%$  below their averages.  This dip event around 8-10 October 1980 is one of the stronger cases for a sudden decrease in brightness over a short period of just a few days.

Two events where KIC 8462852 appears to have flared in the MMO light curve occur on 30 September 1967 and again on 10 August 1977. These events are list in Table~\ref{tab:table_3} and are shown in Figure~\ref{fig:figure_12}. In both cases, the increase in the MMO magnitude is $\sim$10$\%$ above the normal scatter of $\pm$0.07 magnitudes. Three MMO photographic plates were taken on 30 September 1967 (JD 2439764.6, JD 2439764.638, and JD 2439764.694).  Photometry from the first gives $m_{pg}$ = 12.20$\pm$0.09 whereas the next two plates taken in sequence give $m_{pg}$ = 12.56$\pm$0.09 and 12.60$\pm$0.06, 55 minutes and 155 minutes after the first plate, respectively. The next closest MMO photographic plate recorded was three nights before on 27 September 1977, where $m_{pg}$ = 12.50$\pm$0.07.  The Sonneberg light curve does not have data taken near the MMO flare event.  Harvard data taken within 2 days of the MMO flare does not show a flare event. Note that both the Sonneberg and Harvard light curves appear to have corresponding flare events on 30 August -- 1 September 1967, but MMO data does not exist for that date.  

On 10 August 1980, the MMO magnitude was 12.22$\pm$0.06, brighter than the average MMO magnitude of 12.43, and is considered a flare event. Photometry from Harvard plates suggests the the star is gradually increasing from mag = 12.35$\pm$0.11 over 20 days to a peak of 12.20$\pm$0.11 on 10 August 1977, then gradually decreasing for about 15 days to mag = 12.47$\pm$0.16.  The Harvard peak is within 2 days of the MMO flare event. The gradual brightening and fading over $\sim$35 days in the Harvard light curve is reminiscent of the brightening episodes reported by \cite{2017arXiv170807822S}, but in that case lasting several hundred days. The Sonneberg light curve shows an increase in magnitude from 12 August 1977 to 16 August 1977, consistent with the MMO and Harvard data.  The flare events observed from the three different photographic plate collections coincide in 1977 and presents a strong case for a KIC 8462852 flare event.  

\section{Discussion}

The Kepler dip events with dips $\ga10\%$ occurred in March 2011 and about 720 days later in March 2013. Note that the 1978 MMO dip event is followed by another dip event in 1980, about 720 days later. The similarity in timing between the 2011--2013 and 1978--1980 light curve dip events is enticing. Neither the 1935 nor the 1966 dip events appear to have corresponding dip events occurring within 720 days.  The lack of observed dip events separated by 720 days in the 1930's and 1960's may be due to the irregularity in observing times, resulting in gaps in data. However, these times are fairly well observed as seen in the light curve (Figure~\ref{fig:figure_5}), so we might expect to see other dips.  But, if the dips are $\la10\%$, we may not see the dip events in the MMO photographic data where typical uncertainty is $\pm$0.07 mag.  As such, we cannot be certain the 720 days between MMO dip events 1978--1980 is a repetition of the 2011--2013 events.  In any case, 5 apparent dip events were observed between 1922 and 1991 in the MMO light curve.  Models like those of \cite{2018MNRAS.473L..21B} which are based on periodic events cannot be ruled out, and attempts to fit such models to the MMO data may find a period that works.  However, at the same time, debris models like those of  \cite{2017MNRAS.468.4399M}, or models of internal mechanisms \citep[e.g.][]{2016ApJ...829L...3W, 2017ApJ...842L...3F, PhysRevLett.117.261101} may also fit the MMO dip events.  

The 1967 and 1977 MMO flare events occur within one year of the 1966 and 1978 dip events. The same does not occur in 1935 or 1980. Even though 23 MMO photographic plates were taken between July 9 and December 11, 1934, a year before the 1935 MMO dip event, no flare is observed. The same is true one year after, when 36 MMO photographic plates were taken between 3 April and 18 December 1936. Also, a year before the 1980 MMO dip events, 16 MMO photographic plates were taken between 23 April and 22 October 1979, and no flares were observed.  Only 6 MMO photographic plates were taken between 26 April and 8 July 1981, a year after the 1980 MMO dip event, and no flare event was observed.  As such, the flare events cannot be confidently connected to dip events. As pointed out in Section~\ref{sec:section3.2}, the 1977 MMO flare event may be more of a gradual brightening over tens of days.  The 1967 MMO flare lacks data and corresponding data in the Harvard and Sonneberg light curves to say the same.  Without more evidence, we cannot say for certain whether the flares are sudden events or more gradual brightening of the star over tens of days. The 1967 and 1977 data seems to support both claims.

\section{Conclusions}

Our data indicate KIC 8462852 gradually decreased in magnitude over 70 years from 1922 to 1991.  From binned 5-year intervals of the light curve, our derived rate is +0.12 $\pm$0.02 magnitudes/century.  This is consistent with the trend of slow fading first pointed out by \cite{2016ApJ...822L..34S}, who found +0.164 $\pm$ 0.013 from Harvard plates.

Five dip events and two flare events are observed, but the photographic data is too sparse to firmly establish any periodicity to the occurrence of the events. Also, the uncertainty in magnitude is large enough that perhaps other smaller dip or flare events may be lost in the noise.  Constraints on either periodic transit models, debris models, and intrinsic mechanism models depend on future observations.  However, observations from the past 100 years, like those from the MMO photographic plates, complements new observations and provides important tests for the models.  

We point out that the use of archival photographic data to explore events in the light curve of KIC 8462852 can be expanded.  In this paper we started with observed dips and flares in the MMO light curve and then looked for corresponding evidence of similar events from photometry from other photographic plate collections. An interesting study might be to begin with the identification of potential dips and flares in the Harvard collection, for example, and look for corresponding evidence in the Sonneberg and MMO light curves.

\section*{Acknowledgements}

We acknowledge Terry Jednaszewski and Sheldon Kagel for digitizing the photographic plates. We gratefully acknowledge Don Cline and his efforts to secure our astronomical heritage by storing and archiving astronomical photographic plate collections in APDA at the Pisgah Astronomical Research Institute.  We also gratefully acknowledge support from those who contributed to the crowdfunding project called The Astronomy Legacy Project, who helped fund the digitizing equipment that was used in this study. We also acknowledge the countless hours at the Maria Mitchell Observatory telescope by several generations of astronomers who took the photographic plates used in this study.  We thank the anonymous referee for many helpful comments for improvements which helped clarify and focus this paper.




\bibliography{kic8462852.bib} 



\end{document}